\documentstyle[11pt]{article}
\catcode`\@=11

\catcode`\@=12

\newcommand{\cites}[1]{${}{\mbox{\small #1}}$}
\newcommand{\be}{\begin{equation}}
\newcommand{\ee}{\end{equation}}
\newcommand{\ba}{\begin{eqnarray}}
\newcommand{\ea}{\end{eqnarray}}
\newcommand{\bc}{\begin{center}}
\newcommand{\ec}{\end{center}}
\def\ti{\tilde}

\def\vp{\varphi}
\def\ga{\gamma}

\def\nvec#1{\mbox{\bf #1}}

\def\vr{\nvec{r}}

\def\vc{\nvec{c}}

\def\vv{\nvec{v}}
\def\uu{\nvec{u}}
\def\pp{\nvec{p}}

\begin{document}
\title{
 Fock-Lorentz transformations \\
 and time-varying speed of light
}
\author{
S. N. Manida\\
Saint-Petersburg State University\\
{\it e-mail: sman@hq.ff.phys.spbu.ru}}
\date{\today}
\maketitle
\begin{abstract}
The
theory of
 relativity
was built up on
linear Lorentz  transformation.
However, in his fundamental work ``Theory of Space, Time and
Gravitation''
\cites{\cite{fock}}
V.A.Fock
 shows
 that the general form of the transformation
between the coordinates  in the two inertial frames could  be taken to be
linear fractional.  The implicit form  of this  transformation
contains two  constants of different space-time dimensions.
They
can be reduced to the constant $c$ with the dimension of speed
(``speed of light''),
and to
the constant $R$ with  the dimension of  length
 (an invariant radius
of the visible part of the Universe).
The geometry of the ``light cones'' shows that $R$ is a fundamental constant,
but
$c$ depends on the time of transformation.

\vspace{3mm}

 Pacs: 03.30 Special relativity

\begin{flushleft}
NIIF/99-033\\
\end{flushleft}
\end{abstract}

\pagebreak
\subsection{Introduction}

The Theory of Relativity was built up on two postulates, namely,
the principle of relativity
and another principle that states that the velocity of light
is independent of the velocity of its source.
Only the combination of the two principles requires the transformation
formulae
between the orthogonal rectilinear coordinates
in the two inertial frames to be linear.
However, from
1910 it was known
\cites{\cite{von}}, \cites{\cite{von2}}
that constancy of the speed of light
is a consequence of
the principle of relativity and general homogeneity properties
of the space-time
manifold in the case of linear transformations.
Many authors
\cites{\cite{c1}}-\cites{\cite{c14}}
have postulated linear transformations or added  different
(but equivalent)
hypotheses with the purpose to get the constancy of the speed of light
as a consequence of  relativity.

In his fundamental work ``Theory of space, time and gravitation''
\cites{\cite{fock}}
Fock shows that  constancy of the speed of light is equivalent
to the linearity of the transformations.
In the same work in Appendix A it is shown, that
from the principle of relativity alone one can get
a more general {\it linear fractional}
form of the transformation
between two inertial frames (Fock-Lorentz
or FL transformation below).

One can take any conventional method to construct reference frames
(tool kit for space-time measurements).
If the velocity of any body in  system
 $k$ is   constant and equal to $\uu,$
we can introduce  system $k'$ rigidly linked
to the moving body in such a way that the velocity of the origin of
the frame $k$  in $k'$
 is equal to $-\uu.$
The arbitrariness of the common scaling
of  length and time units can be eliminated
by the symmetry of direct and inverse transformations
 (from $k$ to $k'$ and from $k'$ to $k)$ ---
those transformations must differ only by the sign of  $\uu.$

The usual {\it a priori} assumption that all bodies at rest in $k'$ have
 the same velocity $\uu$ in $k$ is equivalent to linearity
of the transformation.
This assumption is not valid in the case of
linear fractional transformations.
So, we had to be careful in the definition
of the relative velocities.
Let us introduce coordinates in such a way that point
$\vr = 0$
at rest
in frame $k$
has constant velocity $\uu$
in $k';$ point $\vr' = 0$
at rest in frame  $k'$ has
constant velocity $-\uu$ in $k.$
We take that  at  $t =  0$  the spatial origins of both frames
coincide and $t'=0.$

Let us write the general
transformation
  for transition from a system $k$ to a system
$k'$
as ratios of linear functions, all
with the same denominator:

\ba
t'=
\frac
{a_{tt}(u)t+a_{tx}(u)x+a_{ty}(u)y+a_{tz}(u)z}
{A(u)+A_{t}(u)t+A_{x}(u)x+A_{y}(u)y+A_{z}(u)z},
\label{F1a}\\
x'=
\frac
{a_{xt}(u)t+a_{xx}(u)x+a_{xy}(u)y+a_{xz}(u)z}
{A(u)+A_{t}(u)t+A_{x}(u)x+A_{y}(u)y+A_{z}(u)z},
 \label{F1b} \\
y'=
\frac
{a_{yt}(u)t+a_{yx}(u)x+a_{yy}(u)y+a_{yz}(u)z}
{A(u)+A_{t}(u)t+A_{x}(u)x+A_{y}(u)y+A_{z}(u)z},
 \label{F1c}   \\
z'=
\frac
{a_{zt}(u)t+a_{zx}(u)x+a_{zy}(u)y+a_{zz}(u)z}
{A(u)+A_{t}(u)t+A_{x}(u)x+A_{y}(u)y+A_{z}(u)z}.
\label{F1d}
\ea
Here $u$ is the relative velocity,
$
a_{\mu \nu}(u),
$
$
A_{\mu}(u)
$
and
$
A(u)
$
are arbitrary functions of the relative velocity.
The denominator in Eqs.(\ref{F1a})-(\ref{F1d})
can go to zero.
If we require the finiteness of coordinates
of all points in all inertial systems,
we simplify Eqs.(\ref{F1a})-(\ref{F1d}) down to the usual
linear transformation:
$ A(u)=1, $ all $ A_{\mu}(u) = 0.  $
This requirement seems natural for uniform Euclidean space.
However,
it became clear
from  Friedmann's work
\cites{\cite{fried}}
that our space-time can have a more complex structure.
Existence of the island masses can lead to  space anomalies,
 uniform
mass distribution across the whole space leads to the big-bang anomalies.
Eqs.(\ref{F1a})-(\ref{F1d}) operates
in such space-time which
can  serve as a background
for the real
Universe scenario.

In Section II we present the derivation of
the explicit form of the FL transformation.

Section III is devoted to some
 geometrical
consequences
of this transformation.
Main of them is the relativity of infinity:
points which are infinitely distant across space-time
in one reference system
appear to be on a
finite distance when viewed from another
system.
  Under the FL transformation, infinity no longer can be
considered as an invariant point.
When
we are going from Galileo to Lorentz transformations,
the infinite invariant
velocity  becomes  finite invariant velocity $c.$
Now, when we are going from Lorentz to Fock-Lorentz
transformations, the infinitely large invariant space-time distance
becomes finite invariant space-time distance.

A new fundamental constant $R$ appears in the FL transformation.
For distances
$
L\ll R
$
and times
$
T\ll R/c,
$
FL transformation becomes
  the usual Lorentz transformation.
 So it is clear that $R$ should
stand for  some cosmological space distance.

In Section IV the
velocity and  the energy-momentum four-vectors are constructed.
They appear to be Lorentz-covariant under
the FL transformation of the space-time manifold.

In Section V  we study
the geometry of
FL-covariant
world lines of
photons, bradions and tachions
and get
an explicit  speed of light  dependence on time:
$
c(t)=R/t,
$
where
$
t=0
$
in the peculiar point of the
transformation in
linear fractions.
In this point all bradions are concentrated into the
invariant
sphere  of radius $R,$
and all photons --- on the surface of this sphere.

In  works
\cites{\cite{l1}}-\cites{\cite{l9}},
 a Varying Speed of Light (VSL) model
is discussed
as a possible alternative to
inflationary cosmology.
This model may resolve some cosmological
puzzels, but the authors usually
postulate  breaking
Lorentz invariance
of space-time or of fundamental interactions.

In our approach, Lorentz invariance is local
but Fock-Lorentz invariance and varying speed of light are
global fundamental properties of  space-time.

In Section VI we show the connection between the
FL transformation and the Friedmann-Lobachevsky
space geometry.

In Section VII we present some formulae
for the red shift
of expanding galaxies and for the usual
longitudial
Doppler effect in the  VSL model.
A non-trivial connection between
the Hubble constant $H_1$ and the red-shift parameter $z$
is established (we write $H_0$ for conventional  Hubble constant $cz/r).$

In Appendix I we explain the mathematical origin
of the linear fractional
transformation
with common denominator.

\subsection{Derivation of the explicit form of the Fock-Lorentz
transformation}

We can rewrite Eqs.(\ref{F1a})-(\ref{F1d})
in vector form and take into account the isotropy
of our space:
\ba
t'=
\frac
{a(u)t+b(u)(\uu\vr)}
{A(u)+B(u)t+D(u)(\uu \vr)},
\label{F2a}\\
\vr'_{||}=
\frac
{d(u)\uu t+e(u)\vr_{||}}
{A(u)+B(u)t+D(u)(\uu \vr)},
 \label{F2b} \\
\vr'_{\perp}=
\frac
{f(u)\vr_{\perp}}
{A(u)+B(u)t+D(u)(\uu \vr)}.
 \label{F2c}
\ea
Here $a(u),$ $b(u),$ $d(u),$ $e(u),$ $f(u),$ $A(u),$ $B(u),$ $D(u)$ are
unknown functions of the relative velocity $u.$

Now we fix these functions.

1.
Transformation
(\ref{F2a})-(\ref{F2c})
will not be changed
after replacement of $\vr_{||}$ for $-\vr_{||},$ $\vr'_{||}$
for $-\vr'_{||},$
$\uu$ for $-\uu,$ since such replacement is equivalent to the
 rotation around
 $\vr_{\perp}$
by the angle $\pi.$

Corollary:  all unknown functions in Eqs.(\ref{F2a})-(\ref{F2c})
are simultaneously odd or simultaneously even.
It is natural to choose even functions.

2. Point $\vr'=0$ has
velocity $\uu$
in frame $k\!:$
$$
d(u)=- e(u)
$$

3. Point $\vr=0$
has velocity $-\uu$
in frame $k'\!:$
$$
a(u)=- d(u) =e(u).
$$

We can substitute
$$
b(u)=-a(u)g(u)
$$

and get
\ba
t'=
\frac
{a(u)(t-g(u)(\uu\vr))}
{A(u)+B(u)t+D(u)(\uu \vr)},
\label{F3a}\\
\vr'_{||}=
\frac
{a(u)(\vr_{||}-\uu t)}
{A(u)+B(u)t+D(u)(\uu \vr)},
 \label{F3b} \\
\vr'_{\perp}=
\frac
{f(u)\vr_{\perp}}
{A(u)+B(u)t+D(u)(\uu \vr)}.
 \label{F3c}
\ea

4. The inverse transformation from $k'$ to
$k$
must be  the same as in Eqs.(\ref{F3a})-(\ref{F3c})
with
$
-\uu
$
instead of
$
\uu\!:
$
\ba
t=
\frac
{a(u)(t'+g(u)(\uu\vr'))}
{A(u)+B(u)t'-D(u)(\uu \vr')},   \nonumber
\\
\vr_{||}=
\frac
{a(u)(\vr'_{||}+\uu t')}
{A(u)+B(u)t'-D(u)(\uu \vr')}, \nonumber
\\
\vr_{\perp}=
\frac
{f(u)\vr'_{\perp}}
{A(u)+B(u)t'-D(u)(\uu \vr')}.  \nonumber
\ea

Combination of direct and inverse
transformations will reduce to the identity,
which means that
\ba
A(u)&=&a(u)\sqrt{1-g(u)u^2},
\nonumber\\
B(u)&=&-\frac{D(u)}{g(u)}\sqrt{1-g(u)u^2},
\nonumber\\
f(u)&=&A(u).
\nonumber
\ea

We can introduce
$$
D(u)=h(u)A(u)
$$
$$
\sqrt{1-g(u)u^2}=\ga^{-1}(u),
$$

and get
\ba
t'=
\frac
{\ga(u)(t-g(u)(\uu\vr))}
{1+\frac{h(u)}{g(u)}\left(1-\ga^{-1}(u)\right)t+h(u)(\uu \vr)},
\label{F5a}\\
\vr'_{||}=
\frac
{\ga(u)(\vr_{||}-\uu t)}
{1+\frac{h(u)}{g(u)}\left(1-\ga^{-1}(u)\right)t+h(u)(\uu \vr)},
 \label{F5b} \\
\vr'_{\perp}=
\frac
{\vr_{\perp}}
{1+\frac{h(u)}{g(u)}\left(1-\ga^{-1}(u)\right)t+h(u)(\uu \vr)}.
 \label{F5c}
\ea

5. Now we write transformation
from $k'$
to other frame
$k'',$
which is moving
with velocity
$\uu'$
in
$k'.$
(We can put
$\uu'||\uu$
without any lack of generality):
\ba
t''=
\frac
{\ga(u')(t'-g(u')(\uu'\vr'))}
{1+\frac{h(u')}{g(u')}\left(1-\ga^{-1}(u')\right)t'+h(u')(\uu' \vr')},
\label{F6a}\\
\vr''_{||}=
\frac
{\ga(u')(\vr'_{||}-\uu' t')}
{1+\frac{h(u')}{g(u')}\left(1-\ga^{-1}(u')\right)t'+h(u')(\uu' \vr')},
 \label{F6b} \\
\vr''_{\perp}=
\frac
{\vr'_{\perp}}
{1+\frac{h(u')}{g(u')}\left(1-\ga^{-1}(u')\right)t'+h(u')(\uu' \vr')},
 \label{F6c}
\ea
Combine Eqs.(\ref{F5a})-(\ref{F5c})
and
(\ref{F6a})-(\ref{F6c}),
and get the transformation from
$k$
to
$k''$
with some new relative velocity
$u''.$
Such transformation,
\ba
t''=
\frac
{\ga(u'')(t-g(u'')(\uu''\vr))}
{1+\frac{h(u'')}{g(u'')}\left(1-\ga^{-1}(u'')\right)t+h(u'')(\uu'' \vr)},
\nonumber
\\
\vr''_{||}=
\frac
{\ga(u'')(\vr_{||}-\uu'' t)}
{1+\frac{h(u'')}{g(u'')}\left(1-\ga^{-1}(u'')\right)t+h(u'')(\uu'' \vr)},
\nonumber
\\
\vr''_{\perp}=
\frac
{\vr_{\perp}}
{1+\frac{h(u'')}{g(u'')}\left(1-\ga^{-1}(u'')\right)t+h(u'')(\uu'' \vr)},
\nonumber
\ea
will take place only if
\ba
1.&& \,\, g(u)=g(u')=g(u'')\equiv \frac{1}{c^2}, \nonumber
\\
2.&&\, \, u''=\frac{u+u'}{1+\frac{uu'}{c^2}},  \nonumber
\\
3.&&\, \,  h(u)=\ga(u)\frac{1}{Rc}.   \nonumber
\ea
Here $c$ is  constant with the  dimension of velocity,
$R$ is constant with the dimension of length.

At last the final form of FL transformation
is
\ba
t'=
\frac
{\ga(u)\left(t-{\uu\vr}/{c^2}\right)}
{1-\left(\ga(u)-1\right)ct/R+\ga(u){\uu \vr}/{Rc}},
\label{F8a}\\
\vr'_{||}=
\frac
{\ga(u)(\vr_{||}-\uu t)}
{1-\left(\ga(u)-1\right)ct/R+\ga(u){\uu \vr}/{Rc}},
 \label{F8b} \\
\vr'_{\perp}=
\frac
{\vr_{\perp}}
{1-\left(\ga(u)-1\right)ct/R+\ga(u){\uu \vr}/{Rc}},
 \label{F8c}
\ea
In the limit
$
r\ll |R|
$
and
$
ct\ll |R|,
$
Eqs.(\ref{F8a})-(\ref{F8c})
coincide with the usual Lorentz transformation.

It is easy to observe that
Eqs.(\ref{F8a})-(\ref{F8c})
are equivalent to the Lorentz transformation
\ba
\frac
{t'}
{1+{ct'}/{R}}
&=&
\frac
{\ga(u)\left(t-{\uu\vr}/{c^2}\right)}
{1+{ct}/{R}},
\label{F10a}\\
\frac{\vr'_{||}}
{1+{ct'}/{R}}
&=&
\frac
{\ga(u)(\vr_{||}-\uu t)}
{1+{ct}/{R}},
 \label{F10b} \\
\frac{\vr'_{\perp}}
{1+{ct'}/{R}}
&=&
\frac
{\vr_{\perp}}
{1+{ct}/{R}},
 \label{F10c}
\ea
for the quantities
\be
\ti{t}=
\frac
{t}
{1+{ct}/{R}},
\qquad  \qquad
\ti{\vr}=
\frac
{\vr}
{1+{ct}/{R}}.
\label{F11}
\ee

\subsection{Properties of the FL transformation}

From Eqs.(\ref{F10a})-(\ref{F11})
one can construct the finite invariant
\be
c^2\ti{t}^2-\ti{\vr}^2=
\frac{c^2t^2-\vr^2}{(1+ct/R)^2}=
\mbox{inv}
\label{F13}
\ee
and the line element
\be
ds^2=c^2d\ti{t}^2-d\ti{\vr}^2 =
\frac{(1-r^2/R^2)c^2dt^2-(1+ct/R)^2d\vr^2+2cdt(1+ct/R)\vr d\vr/R }
{(1+ct/R)^4}.   \label{F14}
\ee
Let us find spacelike and timelike
 ``hyperboloids''. For positive sign of invariant (\ref{F13})
one can get for any world point
\be
\frac{c^2t^2-\vr^2}{(1+ct/R)^2}=\frac{c^2\tau^2}{(1+c\tau/R)^2}
\label{F15}
\ee
where
$
\tau
$ is a ``proper time'' --- time in the space origin of the rest frame.

For
$\tau>0,$
 ``hyperboloid'' (\ref{F15}) has asyptote
$r\rightarrow \infty$
\be
   ct_{as}=\pm r\frac{1+c\tau/R}{\sqrt{1+2c\tau/R}}
+\frac{c^2\tau^2/R}{1+2c\tau/R}.\nonumber
\ee
This ``hyperboloid'' has two parts ---
one in the region
$t>\tau$ and the other in the region $t<-R/c.$

All time-like ``hyperboloids''
have common $S_3^-$ intersection $(ct=-R,$
$\vr^2=R^2)$  of the  light cone $c^2t^2=\vr^2$ with the hyperplane
$ct=-R.$

For negative $\tau$ the entire ``hyperboloid'' is inside the light cone
of the past and includes the same $S_3^-.$

For negative values of interval
(\ref{F13})
we can write
\be
\frac{c^2t^2-\vr^2}{(1+ct/R)^2}=-\lambda^2.
\label{F16}
\ee
Here
$\lambda$
is a space coordinate of the world point in the frame, where
its time coordinate is  $t=0.$
All space-like ``hyperboloids'' are tangent to the light cone
$c^2t^2=\vr^2$ on the same $S_3^-$ manifold.

Asymptotes for $r \rightarrow \infty$ are
\be
   ct_{as}=\pm r\frac{1}{\sqrt{1+\lambda^2/R^2}}
-\frac{\lambda^2/R}{1+\lambda^2/R^2}.  \nonumber
\ee

Assume that some particle was produced
at time $t=0$ and decayed at time $\tau$
in its own rest frame. In another frame of reference, where this particle
had velocity $\uu,$  it decayed at time $t$ in
a point with coordinate
$\vr = \uu t.$
From Eq.(\ref{F13})
we can get its lifetime in the ``moving'' frame as
\be
t=\frac{\tau}{(1+c \tau/R)\sqrt{1-\uu^2/c^2}-c\tau/R}.
\label{F18}
\ee

Some apparent properties of Eq.(\ref{F18}):

A. If the velocity of the particle in any frame is sufficiently great
\be
u>\frac{\sqrt{1+2c\tau/R}}{1+c\tau/R} \label{F18a}
\ee
or, in other words, if its proper lifetime is large enough
$$
\tau >\frac{R/c}{\ga(u)-1},
$$
we get
from Eq.(\ref{F18}) that
the world line of this particle is going from the origin
till infinity  in the future light cone and then from
infinity till  $t<0$ in the past light cone.
So this particle is seen to be ``stable'' in  frames (\ref{F18a}).

 B. For
$\tau=-R/c$
we get from Eq.(\ref{F18})
that this time interval is the same
$(t =-R/c)$ in all inertial frames of reference.
So, in our approach
the special finite time interval $-R/c$ is invariant, but
infinite time intervals are not.

 If any body is at rest in any frame of reference at distance $\vr$
from the origin, then in another frame, moving with  velocity $\uu,$
this body will be at distance $\vr'_0$
from the origin at time $t'=0.$
Let us find
$\vr'_0.$
Put $t' =0,$ $\vr'=\vr'_0$
in Eqs.(\ref{F8a})-(\ref{F8c}), eliminate $t$ and get
\ba
\vr'_{0||}
&=&
\frac
{\ga^{-1}(u)\vr_{||}}
{1+\uu \vr/(Rc)}, \nonumber
\\
\vr'_{0\perp}
&=&
\frac
{\vr_{\perp}}
{1+ \uu \vr/(Rc)}. \nonumber
\ea

Let us take a small body, which is moving
in frame $k$
with velocity
$\uu.$
Space-time coordinates of this body are
$(t,$ $\vr).$
In another frame
$k',$
which is moving with velocity $\vv$ in $k,$
this body has the velocity
$\uu'.$
Let us find
$\uu'.$
At first we can rewrite Eqs.(\ref{F10a})-(\ref{F10c})
for space-time differentials:
\ba
&&\frac
{dt'}
{(1+ct'/R)^2}=
\ga(v)
\frac
{dt\left(1+ \vv\vr/(Rc)\right)-(1+ct/R)\vv d\vr/c^2}
{(1+ct/R)^2},
\label{F22a}\\
&&\frac{d\vr'_{||}(1+ct'/R)- \vr_{||}cdt'/R}
{(1+ct'/R)^2}=
\ga(v)
\frac
{d\vr_{||}(1+ct/R)-cdt(\vv+c\vr_{||}/R)}
{(1+ct/R)^2},
 \label{F22b} \\
&&\frac{d\vr'_{\perp}(1+ct'/R)-\vr'_{\perp}cdt'/R}
{(1+ct'/R)^2}=
\frac
{d\vr_{\perp}(1+ct/R)- \vr_{\perp}cdt/R}
{(1+ct/R)^2}.
 \label{F22c}
\ea
Now we intriduce auxillary quantities
$
\vr_0=\vr-\uu t,
$
and
$
\vr'_0=\vr'-\uu' t',
$
and divide
Eqs.
(\ref{F22b}),
(\ref{F22c})
by
Eq.(\ref{F22a}):
\ba
\uu'_{||}-c \vr'_{0||}/R
&=&
\frac
{\uu_{||}-c\vr_{0||}/R- \vv}
{1-\vv(\uu-c\vr_0/R)/c^2},
\label{F23b}\\
\uu'_{\perp}-c\vr'_{0\perp}/R
&=&
\ga^{-1}(v)
\frac
{\uu_{\perp}-c \vr_{0\perp}/R}
{1-\vv(\uu-c\vr_0/R)/c^2}.
 \label{F23c}
\ea

The latter Eqs.(\ref{F23b}), (\ref{F23c})
are appropriate for the investigation
of the FL transformation properties,
but are not convenient for practical use.
The left hand sides of Eqs.(\ref{F23b}), (\ref{F23c})
 contain  $\uu'$ not only explicitly, but also through $\vr'_0.$
Let us connect $\vr'_0$ and $\vr_0.$
If we rewrite Eqs.(\ref{F10a})-(\ref{F10c})
for the definite world point $t'=0,$
$\vr'=\vr'_0,$  we get
\ba
0&=&
\frac
{\ga(v)\left(t-\vv\vr/c^2\right)}
{1+ct/R},
\label{F24a}\\
\vr'_{0||}
&=&
\frac
{\ga(v)(\vr_{||}-\vv t)}
{1+ct/R},
 \label{F24b} \\
\vr'_{0\perp}
&=&
\frac
{\vr_{\perp}}
{1+ct/R}.
 \label{F24c}
\ea
From Eq.(\ref{F24a})
and definiton $\vr_0=\vr-\uu t,$ we get for this  world point
\be
\vr_0 =\vr -\uu (\vv\vr)/c^2.\label{r0}
\ee
The inverse to  Eq.(\ref{r0}) are
\ba
\vr_{||}
&=&
\frac{\vr_{0||}}
{1-\vv \uu/c^2}, \nonumber
\\
\vr_{\perp}
&=&
\frac
{\vr_{0\perp}+[\vv[\uu \vr_0]]/c^2}
{1-\vv \uu/c^2}. \nonumber
\ea
Now we can put Eqs.(\ref{F24b})-(\ref{F24c}),
in right-hand parts of Eqs.(\ref{F24b}), (\ref{F24c}):
\ba
\vr'_{0||}
&=&
\ga^{-1}(v)
\frac{\vr_{0||}}
{1-\vv(\uu/c^2- \vr_0/(Rc))},\nonumber
\\
\vr'_{0\perp}
&=&
\frac
{\vr_{0\perp}+[\vv[\uu \vr_0]]/c^2}
{1-\vv(\uu/c^2- \vr_0/(Rc))}  \nonumber
\ea
and after some dull calculations
rewrite Eqs.(\ref{F23b})-(\ref{F23c})
in manifest form
\ba
\uu'_{||}
&=&
\frac
{\uu_{||}-\vv-\left(1-\sqrt{1-v^2/c^2}\right)c(\vr_{||}-\uu_{||}t)/R}
{1-\vv(\uu-c(\vr-\uu t)/R)/c^2},
\label{F27b}\\
\uu'_{\perp}
&=&
\frac
{\sqrt{1-v^2/c^2}(\uu_{\perp}-c(\vr_{\perp}-
\uu_{\perp}t)/R)+c(\vr_{\perp}-\uu_{\perp}t+[\vv[\uu\vr]]/c^2)/R}
{1-\vv(\uu-c(\vr-\uu t)/R)/c^2}.
 \label{F27c}
\ea

Inspite of cumbersome appearance, the latter equations are
suitable for  direct calculations --- the velocity $\uu$
in world point $ct,$ $\vr$ (in frame $k$)
is connected
directly
with the velocity $\uu'$ in frame $k'$ in the same world point.

On the cones $\vr= \uu t$ $(\vr_0=0)$
we get the ordinary Einstein law for addition of velocities
\ba
\uu'_{||}
&=&
\frac
{\uu_{||}-\vv}
{1-\vv\uu/c^2},
\label{F27bb}\\
\uu'_{\perp}
&=&
\frac
{\uu_{\perp}\sqrt{1-v^2/c^2}}
{1-\vv\uu/c^2}.
 \label{F27cc}
\ea

 Let us return to the law for addition of velocities in the form
of (\ref{F23b}), (\ref{F23c}).
If in any inertial frame
$k$ some bodies (galaxies below) are moving at time $t=0$
with velocities proportional to the distance from the origin
$\uu =  c \vr_0/R=H_0\vr_0,$
we get in the frame $k'$
$\uu' =
c \vr'_0/R- \vv.$
Transferring  the origin into the other galaxy by the
transformation
$
\vr'_0\rightarrow \vr'_0+R\vv/c
$
gives us
$
\uu' = c \vr'_0/R.
$
It means that the Hubble law of expansion
$(R>0)$
or compression
$(R<0)$
of the Universe
with the constant
$H_0=c/R$
is the same for observers in all galaxies.

If any galaxy $A$ has velocity
$\uu_{A}=c\vr_{0A}/R$ in our frame, we can choose  $\vv$
in Eq.(\ref{F27b})
in such a way that
$$\ga(v)\vv=\uu_A
$$
and get the rest frame of galaxy $A:$ $\uu^{'}_A=0.$
The transformation parameter
$$
  \vv=\frac{\uu_{A}}{\sqrt{1+\uu_{A}^2/c^2}}<c.
$$
is meaningful for all values of the velocity
$\uu_A,$
even for
$\uu_A>c$ !

The experimental value\footnote{It will be shown
in Section VII
that the real value of the Hubble constant
$H_1$  may be,
less than
 $H_0$
in our approach.}
of $H_0$ is
$
H_0=100\cdot h_0\,{\mbox{km}}/{\mbox c}/{\mbox{Mpc}},
\quad \mbox{with}\quad 0.6<h_0<0.8,
$
which means that  FL transformation parameter
$
R=\frac{1}{h_0}3\cdot10^3 {\mbox{Mpc}}.
$

In some works
 \cites{\cite{h1}}-\cites{\cite{h3}}
an original approach to the theory of relativity
was introduced with universal constant
$H_0.$
In our approach it is possible, in general, to choose any pair
from
$R,$ $c,$ $H$
as universal
constants.
But, as it will be seen in Section V,
the speed of light at the instant  $t$
from the special point of FL transformation
is equal to $c=R/t.$
So, $R$ is really a fundamental constant, but
 $c=R/t$ and $H=1/t$ are in some sense ``time-measurement''
quantities.

\subsection{Energy-momentum four-vector}

Let us write the metric line element  (\ref{F14}) as
\be
d{s}^2=
\frac{c^2dt^2\left(1-\left({\vv}/{c}+
{\vv t}/{R}-{\vr}/{R}\right)^2\right)}
{\left(1+{ct}/{R}\right)^4}.
\label{F29}
\ee
or, for  $ds^2>0,$  as
\be
ds=cdt
\frac{\sqrt{1-\left({\vv}/{c}-{\vr_{0p}}/{R}\right)^2}}
{(1+ct/R)^2},
\label{F30}
\ee
where
$\vv=d\vr/dt$  is the instantaneous velocity of some particle,
   $ \vr_{0p}=\vr-\vv t $ is an auxiliary ``initial''
   space coordinate of the particle (the coordinate of the particle
   at the instant $t=0$ in the case of a constant velocity $\vv).$

   Now we can introduce the four-vector
   \be
   dx^{\mu}=\frac{1}{(1+ct/R)^2}\left(cdt,d\vr(1+ct/R)-\vr cdt/R\right),
\label{F31}
\ee
   which is constructed from the differentials of
 the Eq.(\ref{F11})
   and  undergoes the usual Lorentz transformation
   during the FL transformation of space-time.
From Eqs.(\ref{F30}), (\ref{F31})
   it is possible to construct the velocity four-vector:
\be
   u^{\mu}=\frac{dx^{\mu}}{ds}=
\frac{1}
{\sqrt{1-\left({\vv}/{c}-{\vr_{0p}}/{R}\right)^2}}
\left(1,{\vv}/{c}-{\vr_{0p}}/{R}\right),
\label{F32}
\ee
   and the energy-momentum four-vector
\be
   p^{\mu}=mu^{\mu}=
\frac{m}
{\sqrt{1-\left({\vv}/{c}-{\vr_{0p}}/{R}\right)^2}}
\left(1,{\vv}/{c}-{\vr_{0p}}/{R}\right).
\label{F33}
\ee
   We shall name the quantities in the right-hand side
   of Eq.(\ref{F33})
as energy
\be
   E=
\frac{m}
{\sqrt{1-\left({\vv}/{c}-{\vr_{0p}}/{R}\right)^2}}
\label{F34}
\ee
   and linear momentum
\be
   \pp=\frac{m\left({\vv}/{c}-{\vr_{0p}}/{R}\right)}
{\sqrt{1-\left({\vv}/{c}-{\vr_{0p}}/{R}\right)^2}}
\label{F35}
\ee
of the particle with mass $m.$
The four-vectors of Eqs.(\ref{F32}), (\ref{F33})
transform like usual Lorentz vectors.

The energy and momentum get minimum values
 $ E_0 = m $ and
$
\pp_0 = 0
$
not on the manifold of the  rest frames  (with
$
\vv = 0),
$
connected by translations,
but on the manifold of the ``Hubble'' frames (with
$
\vv = c\vr_{0p}/R).
$

FL transformations introduce corrections to the
nonrelativistic dynamics
$
(|\vv|\ll c),
$
which are significant at large distances
$
|\vr_{0p}|\approx R|v|/c.
$
For the velocity  1 m/s this distance is near 30 l.y.

Corrections to the relativistic dynamics are significant
at extremly large distances
$
|\vr_{0p}|\approx R
$
only.
Nevertheless, we see from Eqs.(\ref{F34}), (\ref{F35})
that even for
$
|\vr_{0p}|\ll R
$
a particle can move ``faster'' than light\footnote{It is useful to note
that the velocity of light at point $\vr$ depends on its direction
and varies from $c\left(1-r_{0p}/R\right)$
to $c\left(1+r_{0p}/R\right).$ The velocity of the particle at some point
may be more than $c,$ but less than the velocity of light at this point
in the same direction.}
if $c<|\vv|<c(1+r_{0p}/R).$

\subsection{Photons, bradyons, tachyons}

One can see from Eqs.(\ref{F34}), (\ref{F35}), that
for ``bradyons'' (particles on the timelike
worldlines with $m^2>0$)
the velocities and ``initial''  coordinates
satisfy an inequality
$|\vv/c - \vr_{0p}/R| < 1.$
For ``tachyons'' $(m^2<0)$
the relevant inequality is $|\vv/c - \vr_{0p}/R| >1,$
and  ``photons'' $(m^2 = 0)$
are moving along the ``light cones'' $|\vv/c -
\vr_{0p}/R| =1.$

Let us examine the geometry of ``light cones'' more carefully.
The line element (\ref{F29}) is
 invariant under FL transformations
(\ref{F8a})-(\ref{F8c}).
In these transformations the origins of coordinates coincide
at the instant
$
t = t' = 0.
$
The special invariant point of transformations
(\ref{F8a})-(\ref{F8c}) is
$
t = t' =- R/c.
$
Let us change the time scale
$
t\Rightarrow t - R/c,$
$
t'\Rightarrow
t' - R/c.
$
 Now the origins of space coordinates coincide at the time
$
t =t' = R/c\equiv t_0.
$
We denote the velocity of  $k'$ in $k$
by
$
\uu\equiv -\vr_0/t_0,
$
 where
$
\vr_0 = -\uu t_0
$
is the origin of coordinates of $k'$ in $k$ at instant
$
t = 0.
$
Eqs.(\ref{F8a})-(\ref{F8c}) may be written as
\ba
t'=
\frac
{t}
{\ga_0-\left(\ga_0-1\right)t/t_0-\ga_0
{\vr_0 \vr}/{R^2}},
\label{F37a}\\
\vr'_{||}=
\frac
{\ga_0\left(\vr_{||}+\vr_0{t}/{t_0}-\vr_0\right)}
{\ga_0-\left(\ga_0-1\right)t/t_0-\ga_0
{\vr_0 \vr}/{R^2}},
 \label{F37b} \\
\vr'_{\perp}=
\frac
{\vr_{\perp}}
{\ga_0-\left(\ga_0-1\right)t/t_0-\ga_0
{\vr_0 \vr}/{R^2}},
 \label{F37c}
\ea
$(\ga_0\equiv 1/\sqrt{1-r_{0}^2/R^2},$ $r_0<R$ if $u<c).$

The metric line element (\ref{F29}) takes the form
\be
d\hat{s}^2=
\frac{t_0^2dt^2}
{t^4}\left(R^2-(\vr-\vv t)^2\right)=
\frac{t_0^2dt^2}
{t^4}\left(R^2-r_{0p}^2\right).
\label{F38}
\ee
Here
$
\vr_{0p}\equiv \vr-\vv t
$
 is an auxiliary space coordinate of the particle
   at the special point $t=0$ in the case of a constant velocity
$\vv.$

Now the velocity four-vector (\ref{F31}) is
\be
   dx^{\mu}=
\frac{t_0dt}{t^2}\left(R,\vv t-\vr \right)=
\frac{t_0dt}{t^2}\left(R,-\vr_{0p} \right),
\label{F39}
\ee
and the energy-momentum four-vector
(\ref{F33}) is
$\left(\ga_{0p}\equiv 1/\sqrt{1-r_{0p}^2/R^2}\right)$
\be
   p^{\mu}=
m\ga_{0p}\left(1,\frac{\vv t -\vr}{R}\right)=
m\ga_{0p}\left(1,\frac{ -\vr_{0p}}{R}\right).
\label{F40}
\ee

Energy and momentum of the particle with mass $m$ are now
\be
   E=
\ga_{0p} m=\frac{m}{\sqrt{1-(\vr-\vv t)^2/R^2}},
\label{F41}
\ee
\be
   \pp=\ga_{0p}m\frac{-\vr_{0p}}{R}=
m\frac{\vv t - \vr}{R\sqrt{1-(\vr -\vv t)^2/R^2}}.
\label{F42}
\ee
At the instant
$
t = 0
$
 all bradyons are inside the sphere
$
\vr^2=R^2
$
and their  energy and momentum are connected with their position
by  the relation
\be
\frac{\vr_{0p}^2}{R^2}=1-\frac{m^2}{E^2}=
\frac{\pp^2}{E^2}   \label{F43}
\ee
Their velocity is arbitrary and does not depend on energy or momentum.

In the case of a lightlike invariant (\ref{F38})
$
d\hat{s}^2=0
$
we see that all world lines of photons $(m^2=0)$ are passing
through the points
$
r_{0p}^2 = R^2
$
at $t=0$.

At those points their velocities were arbitrary.
Having velocity
$
\vc
$
at $t=0,$
photons are moving along the world line
\be
 \vr(t,c)=\vr_{0p}+\vc t.\label{F45}
\ee
The velocity of the photons in any world point
$t,$ $\vr$
is equal to
\be
 \vc(t,\vr)=\frac{\vr-\vr_{0p}}{t}.\label{F46}
\ee
The photons, which pass through the origin of
the space coordinates
$\vr = 0$
at the time
$t,$
have the velocity
\be
 \vc(t)=\frac{-\vr_{0p}}{t},
\quad \quad
 c(t)=\frac{R}{t}.
\label{F47}
\ee
 The light cone with the vertex at the world point
$t_1,$
$r_1$
is determined by the relation
     \be
 \vr=\vr_{0p}+t\frac{\vr_1-\vr_{0p}}{t_1}, \qquad r_{0p}^2=R^2.\label{F48}
\ee

The dependence of the speed of light on time and space coordinates
does not contradict the constancy of $c$
in the FL transformation
(\ref{F37a})-(\ref{F37c}).
This transformation relates the reference frames
with coincident space  origins
at time
$
t_0.
$
 The constant $c$ in this transformation is equal
to the speed of light at world point
$t_0,$ $0,$  which is $c=R/t_0.$
FL transformations at other moments
$t$ will have other constant $c=R/t.$
The metric line element (\ref{F38}) and four-vector (\ref{F39})
depend on $t_0,$
but four-vector (\ref{F40}) and all subsequent
ones are independent on $t_0.$
Eqs.(\ref{F34}), (\ref{F35}) for energy and momentum
contain $c$ as though they depend on the time of the FL transformation.
Nevetheless, Eqs.(\ref{F41}), (\ref{F42})  obviously demonstrate,
that  energy and   momentum
of the particle are completely determined by the
straight world line $\vr=\vr_{0p}+\vv t,$
which is tangential to the world line of
this particle.

The quantity
 \be
ds^2=\frac{d\hat{s}^2}{t_0^2}=\frac{dt^2}{t^4}(R^2-r_{0p}^2)
\label{F49}
\ee
 is invariant under FL transformations
(\ref{F37a})-(\ref{F37c})
for any values of
$t_0$ ¨ $\vr_0.$\footnote{Let us notice
that
$\vr_0$ determines
the velocity $\uu$ of
$k'$ in
$k\!:$
$\uu =
-\vr_0/t_0,
$
and
$
\vr_{0p}
$
determines the velocity \vv of the particle in
$k,$ i.~e.
$
\vv = (\vr-\vr_{0p})/t.
$}

We have introduced  energy $E$ and momentum $\pp$ by
Eqs.(\ref{F41}), (\ref{F42}).
They differ from the current quantities $\hat{E},$
$\hat{\pp}$ by the dimensional multiplier:
\be
   \hat{E}=
\ga_{0p} mc^2=
\ga_{0p} m\frac{R^2}{t^2},
\label{F50}
\ee
\be
   \hat{\pp}=
\ga_{0p}mc\frac{-\vr_{0p}}{R}=
\ga_{0p}m\frac{-\vr_{0p}}{t}.
\label{F51}
\ee
In our approach these quantities are not conserved in time.
Apparently, in the dynamics with varying speed of light
just $E$ and $\pp$ from Eqs.(\ref{F41}) and
(\ref{F42}) should be conserved.

It is interesting to note that in the limit
$
t_0\rightarrow 0,
$
taking into account
$
\vr_0 = -\uu t_0
$
and
$
\gamma_0\rightarrow 1+(\uu t_0)^2/(2R^2),
$
we  get Galilean transformation from
Eqs.(\ref{F37a})-(\ref{F37c})\footnote{In article
\cites{\cite{Liu}}
some modification of the special theory of relativity
was suggested in which
 auxiliary Galilean coordinate systems  were introduced
for every inertial frame of reference.
It induces the localized Lorentz transformation between any
two usual inertial coordinate systems.}:
\be
t'=t, \quad \quad  \vr'=\vr-\uu t.
\label{F52}
\ee
Indeed, the speed of light is going to infinity in the limit
 $
t_0 \rightarrow 0.
$

Let us note also that the sphere
$
\vr^2\leq R^2
$
 at
$
t=0$
is invariant
under FL transformation
(\ref{F37a})-(\ref{F37c}).
We can put
 for clarity
$R=1$ and get from
Eqs.(\ref{F37a})-(\ref{F37c})
for $t=0$
\ba
t'&=&0, \label{85}\\
\vr'_{||}&=&
\frac{\vr_{||}-\vr_0}
{1-\vr_0\vr},
 \label{86}\\
\vr'_{\perp}&=&
\frac{\vr_{\perp}\sqrt{1-r_0^2}}
{1-\vr_0\vr}.
 \label{87}
\ea
This transformation is equivalent to
\ba
1-\vr'^2&=&
(1-\vr^2)
\frac{1-\vr_0^2}
{(1-\vr_0\vr)^2},
 \label{88}\\
\frac{1-\vr'^2}{1+\vr_0\vr'}&=&
\frac{1-\vr^2}
{1-\vr_0 \vr}.
 \label{89}
\ea
In new variables
\ba
&&z=1-\vr^2, \nonumber \\
&&z'=1-\vr'^2,\nonumber \\
&&w=1-\vr\vr_0,\nonumber \\
&&w'=1+\vr'\vr_0,  \nonumber
\ea
which are connected by relations
$ ww'=(1-r^2_0),$
$z'/w'=z/w,$
the elementary volume transforms as
      $$
dV'=dV\frac{(1-r_0^2)^2}{w^4}
$$
or
$$
\frac{dV}{z^2}=\frac{dV'}{z'^2}.
$$
So
$$
f(r)=\frac{f(0)}{(1-\vr^2)^2}
$$
is an invariant distribution function
inside the sphere
$
\vr^2\leq 1
$
at $t=0.$

\subsection{Connection with the
Friedmann-Lobachevsky metric}

Let us rewrite the metric line element (\ref{F38}) as
\be
d\hat{s}^2= \frac{t_0^2}{t^4}
\{R^2dt^2-(\vr dt-t d\vr)^2\}, \label{Ff}
\ee
and single out the full square with $dt:$
\be
d\hat{s}^2=
\frac{t_0^2}{t^4}\left\{\left[\sqrt{R^2-r^2}dt+
\frac{tr dr}{\sqrt{R^2-r^2}}
\right]^2
-\frac{t^2}{t_0^2}\left[\frac{R^2dr^2}{R^2-r^2}+r^2(d\theta^2+\sin^2\theta
d\vp^2)\right]\right\}.
\label{F71}
\ee
After the replacement
\be
t=\hat{t}\sqrt{1-\frac{r^2}{R^2}},\label{Fm}
\ee
which differs from the known Milne transformation
\cites{\cite{Milne}}
\be
t=\hat{t}\sqrt{1-\frac{u^2}{c^2}}, \label{Mln}
\ee
by the region of definition\footnote{
Eq.(\ref{Mln}) is definite
in the future light cone and Eq.(\ref{Fm}) is definite
in all observable space-time cylinder $r<R.$},
we get the line element
\be
d\hat{s}^2=
\frac{R^2t_0^2}{\hat{t}^4}\left(d\hat{t}^2
-\frac{\hat{t}^2}{R^2} d{\ell}^2\right)\\,\label{F72}
\ee
and curved space-like hypersection with the metric
\be
d\ell^2=\frac{1}{1-r^2/R^2}\left(\frac{dr^2}
{1-r^2/R^2}+r^2(d\theta^2+\sin^2\theta
d\vp^2)\right).
\label{F73}
\ee
After standard replacements
$$
r=\frac{\tilde{r}}{\sqrt{1+\tilde{r}^2/R^2}},
$$
or
$$
r=\frac{\hat{r}}{1+\hat{r}^2/(4R^2)},
$$
we get well known forms of Friedmann-Lobachevsky space
\be
d\ell^2
=\frac{d\tilde{r}^2}{1+\tilde{r}^2/R^2}+
\tilde{r}^2(d\theta^2+\sin^2\theta d\vp^2)
=\frac{d\hat{r}^2+\hat{r}^2(d\theta^2+\sin^2\theta d\vp^2)}
{\left(1+\hat{r}^2/(4R^2)\right)^2}.
\label{F74}
\ee

  Eqs.(\ref{F72}), (\ref{F74})  correspond to an open
(space curvature $K=-1/R^2)$ flat expanding Universe
with linear scale factor $\hat{t}/t_0$
in conformal coordinates.
In the work
\cites{\cite{Kh}} it was shown that linear dependence
of the scale factor on time is not in contradiction with
modern astronomical observations
\cites{\cite{SN1}},
\cites{\cite{SN2}}.
Nevertheless, the comparison of the above results
with  experiment would be possible only after
investigation of the red-shift parameters. The varying speed of light
may change many usual relations.

\subsection{Red shift}

Let us investigate the red shift of the light signal from the emitter
in galaxy $A,$ which has constant velocity $\uu$
relative to our Galaxy. We put the space coordinate origin into our
Galaxy and put $t=0$ at the special point of FL transformations
(with the infinite
speed of light).
Let the visual big-bang instant equal to $t_{\ast},$
which may differ from
$t=0.$

We observe at $t_1$ the signal, emitted in galaxy $A$
at $t_2,$ when it  was at distance $\vr_2=(t_2-t_{\ast})\uu$ from us.
The velocity of this signal is $|c_1|=R/t_1.$
The moments $t_1$ and $t_2$ are related by
$$
(t_2-t_{\ast})u=(t_1-t_2)\frac{R}{t_1},
$$
or
\be
  t_2=\frac{R+ut_{\ast}}{R+ut_1}t_1 \label{Ft2}
\ee
and
\be
  \frac{dt_2}{dt_1}=\frac{R(R+ut_{\ast})}{(R+ut_1)^2}.\label{Ft1}
\ee
Let us connect $t_2$ with the emitter proper time $\tau_2.$
From the invariance of interval (\ref{Ff}) between the
world points $(t_2,$ $\vr_2)$ and
$(t_2+dt_2,$ $\vr_2+\uu dt_2),$ we get
\be
\frac{d\tau_2}{\tau_2^2}\gamma=\frac{dt_2}{t_2^2}, \label{F77}
\ee
where
$$
\gamma=\frac{1}{\sqrt{1-u^2/c_{\ast}^2}},
$$
and $c_{\ast}=R/t_{\ast}$
is the speed of light at big bang\footnote{The values of  $t_{\ast}$  and
$c_{\ast}$ may be negative.  This would mean that big-bang time $t_{\ast}$
preceeds the special point $t=0$ of the FL transformation.  At this point
galaxy $A$ would be at distance $r_0=-ut_{\ast}$ from our Galaxy.}.

After integrating of Eq.(\ref{F77})
from $t_{\ast}$ till $t_2$
or directly
from the FL transformation Eqs.(\ref{F37a}), (\ref{F37b}),
we get the relation
\be
\frac{t_2}{\tau_2}=\left(1-\gamma^{-1}\right)
\frac{t_2}{t_{\ast}}+\gamma^{-1}
\label{F78}
\ee
and with the help of Eq.(\ref{Ft2})
\be
\frac{t_2}{\tau_2}=\left(1-\gamma^{-1}\right)
\frac{R+ut_{\ast}}{R+ut_1}\frac{t_1}{t_{\ast}}+\gamma^{-1}
\label{F79}
\ee
From Eqs.(\ref{Ft1})-(\ref{F79})
 we get
\be
z+1= \frac{dt_1}{d\tau_2}=
\left[1+(\gamma-1)
\frac{R+ut_{\ast}}{R+ut_1}\frac{t_1}{t_{\ast}}\right]^2
\frac{(R+ut_1)^2\sqrt{R^2-u^2t_{\ast}^2}}{R(R+ut_{\ast})}.
\label{F80}
\ee

A more evident form of  relation (\ref{F80})
may be written with the help of $c_1$ and $c_{\ast}$
\be
z+1=
\left(1+\frac{u}{c_1}\right)^2\frac{\sqrt{1-u/c_{\ast}}}{\sqrt{1+u/c_{\ast}}}
\left[1+
\left(\gamma-1\right)
\frac{u+c_{\ast}}{u+c_1}\right]^2.
\label{F81}
\ee
In the small $z$ limit
$
(u\ll c_1,$ $u< |c_{\ast}|)
$
we get
$$
z\approx \frac{u}{c_1}\left(2-\frac{c_1}{c_{\ast}}\right).
$$

In the limit $|c_{\ast}|\rightarrow \infty$
$
(|c_{\ast}|\gg c_1,
$ $
|c_{\ast}|\gg u,
$
with any relation between $u$ and $c_1)$
Eq.(\ref{F81}) simplifies:
\be
z+1=
\left(1+\frac{u}{c_1}\right)^2,
\label{F82}
\ee

Independently from the movement of the galaxies we can get the general
form of the Doppler red shift  if we change in Eq.(\ref{F81})
the speed of light at the big bang $c_{\ast}$
by the relation
  \be
  \frac{1}{c_{\ast}}=\frac{1}{c_1}-\frac{r_2}{R}\left(\frac{1}{c_1}+
\frac{1}{u}\right).     \label{c0c1}
\ee
The last equation can be derived from
$
R=c_1t_1=c_{\ast}t_{\ast},
$
$
r_2=c_1(t_1-t_2)=u(t_2-t_{\ast}).
$
Here positive values of $u$
mean the movement of the emitter from the observer
and negative $u$ --- the movement to the observer.

  Now, with the help of Eq.(\ref{c0c1}), we get instead of Eq.(\ref{F81})
\be
z+1=
\frac{\sqrt{1+\frac{u}{c_1}}
\left\{\frac{u}{c_1}
\sqrt{1-\frac{r_2}{R}}-\frac{r_2}{R}
\sqrt{\left(1+\frac{u}{c_1}\right)\left[1-\frac{u}{c_1}
+\left(1+\frac{u}{c_1}\right)\frac{r_2}{R}\right]}\right\}^2}
{\left[\frac{u}{c_1}-\left(1+\frac{u}{c_1}\right)\frac{r_2}{R}\right]^2
\sqrt{\left(1-\frac{r_2}{R}\right)
\left[1-\frac{u}{c_1}+\left(1+\frac{u}{c_1}\right)\frac{r_2}{R}\right]}}.
\nonumber
\ee

Let us investigate sime limiting cases.

1. The longitudinal Doppler effect
$
(r_2\ll R)\!:
$
\be
z+1=\sqrt{\frac{c_1+u}{c_1-u}}, \label{F831}
\ee
which is the usual form.

2. Ultrarelativistic emitter, moving to the observer,
$
(u\approx -c_1)\!:
$
\be
z+1=\sqrt{\frac{R-r_2}{2R}}\sqrt{1+\frac{u}{c_1}}
\label{F832}
\ee

Here we have
blue shift, which is enhananced for the emission from the
region
$
r_2 \rightarrow R.
$

3. Emitter with any velocity $u \neq -c_1$  in the region
$
r_2 \rightarrow R$
\be
z+1=
\sqrt
{\frac{2R}
{R-r_2}}
\left(1+\frac{u}{c_1}\right)^{3/2}
\label{F833}
\ee

For all velocities $u\neq -c_1$
we get ``geometrical'' red shift
from the region
$
r_2 \approx R,
$
due to the denominator in Eq.(\ref{F833})

4. For emitter at rest $(u=0)\!:$
\be
z+1=\sqrt{\frac{R+r_2}{R-r_2}}.
\label{F834}
\ee
We get ``distant'' red shift, which has quasi-Hubble
form
$
z\approx {r_2}/{R}
$
for small
distances
$
r_2\ll R.
$

5. The first order contributions of the velocity of the emitter
$
u\ll c_1
$
and  of the distance from it
$
r_2\ll R\!
$
to the red shift value are
\be
z= \frac{u}{c_1}+\frac{r_2}{R}.
\label{F835}
\ee

Let us connect this result with the Hubble constant
$
H_1=u/r_1.
$
Here
$
r_1
$
is the modern distance to the galaxy, which we ``see''
at distance $r_2.$
From the obviouse relations
$
r_1-r_2=u(t_1-t_2)
$
and
$
r_2=c_1(t_1-t_2)
$
 we get
$
r_1=r_2(1+u/c_1)
$
and
\be
  H_1=\frac{uc_1}{r_2(c_1+u)}.  \label{Fh}
\ee
For small
$z$
we have
$
H_1\approx u/r_2
$
and from Eq.(\ref{F835})
\be
   H_1=\frac{zc_1}{r_2}-\frac{c_1}{R}.
\ee

We see that in our approach
the Hubble constant is less than $H_0=zc_1/r_2$ in the Standard Model
of expanding Universe.
It appears impossible to get the Hubble constant directly from
its $z$ dependence on $r_2$ for nearest galaxies.
We need some further assumptions. For example, if
$
|c_{\ast}|\gg c_1
$
we have
$
H_1\approx c_1/R
$
or
$$
H_1=\frac{zc_1}{2r_2}\approx 30 \,
\mbox{km}/\mbox{c}/\mbox{Mpc},
$$
which is two times less than in the Standard Model.
The corresponding age of the Universe
$$
T=\frac{R}{c_1}=H_1^{-1}\approx 33\cdot 10^9\, \mbox{y}
$$
is two times more than the standard one.

\subsection{Conclusions}

We have shown that on the basis of the principle of relativity
it is possible to get a more general geometry of
matter free  space-time
in which the speed of light depends on the time of observation.
A new constant $R$ (an invariant radius of the visible part of the
Universe) appears in this approach.

This model may be seen as an alternative tool for the interpretation of
different cosmological events.

The Newtonian world with the infinite value of $c$ is a limiting
point of the relativistic world. On the same basis the Einstein world
with infinite value of $R$ may be seen as a limiting point in
more general approach.

If the real value of $R$ in our world is  infinitly large and
the speed of light does not depend on time, we need to find some physical
principle which rules out such variant of the space-time structure.

\vspace{0.5cm}
{\Large \bf Appendix A}
\vspace{0.3cm}

After this paper was completed an interesting work
\cites{\cite{p02}} was published.
In this work the authors
start from the inertial frame with global coordinate system $(x^0,$ $x^1,$
$x^2,$ $x^3)$
and look for all coordinate systems
$\xi^{\mu}(x,v),$
such that
\be
\frac{d^2\xi^{\mu}}{ds^2}=0.  \label{A00}
\ee
In view of
$dx^{\mu}/ds=v^{\mu},$
and
$dv^{\mu}/ds=0$ they get
\be
\frac{\partial^2\xi^{\mu}}{\partial x^{\alpha}\partial x^{\beta}}
v^{\alpha}v^{\beta}=0,   \label{A01}
\ee
{\it i.e.} $\xi^{\mu}$ must be linear in $x^{\mu}.$
But Eq.(\ref{A00}) is redundant for the free motion
of the particle. We can put
$$
\frac{d}{ds}\left(\frac{d\xi^{\mu}}{ds}/\frac{d\xi^{0}}{ds}\right)=0
$$
instead of
(\ref{A00}).
After some complicated calculations
(all details can be seen in \cites{\cite{fock}})
we find
$$
\frac{\partial^2(w\xi^{\mu})}{\partial x^{\alpha}\partial x^{\beta}}
v^{\alpha}v^{\beta}=0
$$
instead of (\ref{A01}), where
$$
\frac{\partial^2w}{\partial x^{\alpha}\partial x^{\beta}}
v^{\alpha}v^{\beta}=0.
$$
So, $w$ must be linear,
 but all four functions $\xi^{\mu}$ may be linear fractions in
$x^{\nu}$  with equal denominators $w,$ as we postulated in
Eqs.(\ref{F1a})-(\ref{F1d}).  Only if $w$ is constant, $\xi^{\mu}$
are reduced to linear functions.

\end{document}